\documentclass[twocolumn]{aastex7}

\usepackage{bm,amsmath}

\shorttitle{Near-surface shear layer in calculation}
\shortauthors{H. Hotta}

\graphicspath{{./}{figures/}}
\newcommand{\figpath}{.}
\begin{document}
  \title{Simultaneous construction of fast equator, poleward meridional flow, and near-surface shear layer in solar magnetohydrodynamic calculation}

  \author[0000-0002-6312-7944]{H. Hotta}
  \affiliation{Institute for Space-Earth Environmental Research, Nagoya University \\
  Furo-cho, Chikusa-ku, Nagoya, Aichi 464-8601, Japan
  }
  \email{hotta.h@isee.nagoya-u.ac.jp}
  \begin{abstract}
    We carry out an unprecedented high-resolution simulation for the solar convection zone. Our calculation reproduces the fast equator and near-surface shear layer (NSSL) of differential rotation and the near-surface poleward meridional flow simultaneously. The NSSL is located in a complex layer where the spatial and time scales of thermal convection are significantly small compared with the deep convection zone. While there have been several attempts to reproduce the NSSL in numerical simulation, the results are still far from reality. In this study, we succeed in reproducing an NSSL in our new calculation. Our analyses lead to a deeper understanding of the construction mechanism of the NSSL, which is summarized as: 1) rotationally unconstrained convection in the near-surface layer transports the angular momentum radially inward; 2) sheared poleward meridional flow around the top boundary is constructed; 3) the shear causes a positive kinetic $\langle v'_r v'_\theta\rangle$ and negative magnetic $\langle B_r B_\theta\rangle$ correlations; and 4) the turbulent viscosity and magnetic tension are latitudinally balanced with the Coriolis force in the NSSL. We emphasize the importance of the magnetic field in the solar convection zone.
  \end{abstract}

  \keywords{Solar convection zone (1998) --- Solar differential rotation (1996) --- Solar dynamo (2001) --- Solar magnetic fields (1503)}

  
  \section{Introduction}
  \label{sec:introduction}
\begin{figure}[htp]
    \centering
    \includegraphics[width=0.4\textwidth]{\figpath/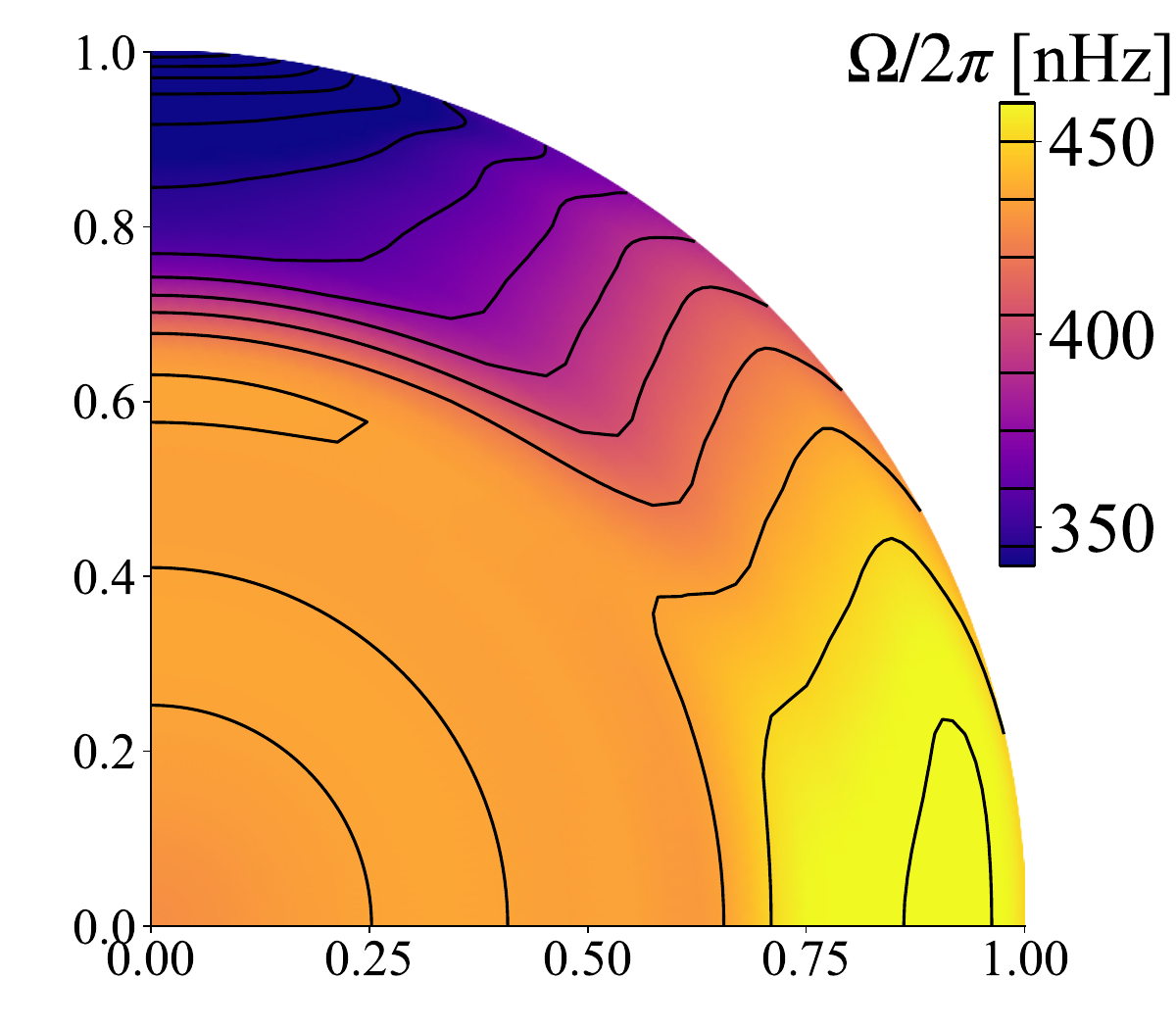}
    \caption{Inversion of the helioseismic data from the SDO (Solar Dynamics Observatory)  satellite for the angular velocity ($\Omega/2\pi$) in the unit of nHz \citep{howe_2011JPhCS.271a2061H}. 
    The solid lines show the values from 340 to 460 nHz in 10 nHz increments.
    The horizontal and vertical axes show the fractional radii.
    \label{difrot_observe}}
  \end{figure}

  The Sun is rotating differentially with the fast equator and the slow pole. Helioseismology has revealed the detailed distribution of the angular velocity $\Omega$ in the solar interior. Fig. \ref{difrot_observe} shows one of the helioseismic results of the differential rotation \citep{howe_2011JPhCS.271a2061H}.
  
  In the solar convection zone, we have two shear layers, i.e., the tachocline around the base of the convection zone and the near-surface shear layer (NSSL). The tachocline is thought to be maintained by the interaction between the convection and radiation zones \citep{spiegel_1992A&A...265..106S,gough_1998Natur.394..755G,forgacs-dajka_2001SoPh..203..195F,Rempel_2005ApJ...622.1320R,brun_2011ApJ...742...79B,matilsky_2022ApJ...940L..50M}. The NSSL is thought to be maintained by the small-spatial and short time scales of the convection in the layer.  The convection spatial scale is roughly scaled with the pressure scale height $H_p\sim T/g$, where $T$ and $g$ are the temperature and the gravitational acceleration, respectively. In addition, the mixing length theory predicts that the convection velocity $v_\mathrm{c}$ varies with the density $\rho$ as $v_\mathrm{c}\sim \rho^{-1/3}$. The pressure scale height at $r=0.71R_\odot$ and $0.99R_\odot$ is 60 and 2 Mm, respectively. The densities in these layers are 0.2 and $3\times10^{-4}~\mathrm{g~cm^{-3}}$, respectively. Thus, the time scales of the convection range from a month to several hours in these regions. As a result, the convection in the NSSL is not significantly affected by the rotation. \cite{foukal_1975ApJ...199L..71F} suggest that the short time scale convection causes negative correlation $\langle v'_r v'_\phi\rangle$ and leads to the radially inward angular momentum transport, where $v_r$ and $v_\phi$ are the radial and longitudinal velocities, respectively. $\langle\rangle$ and $'$ denote the longitudinal average and the deviation from the average. 
  In addition, \cite{Miesch_2011ApJ...743...79M} suggest that we need a force to balance with the latitudinal Coriolis force to maintain the NSSL.
  It is difficult for numerical simulations to cover a broad range of spatial and time scales. The numerical approach for the NSSL is highly restricted. \cite{guerrero_2013ApJ...779..176G} enhance the superadiabaticity around the top boundary of their calculation box and discuss the formation mechanism of the NSSL following \cite{foukal_1975ApJ...199L..71F}. \cite{Hotta_2015ApJ...798...51H} for the first time, cover the convection zone from $0.71R_\odot$ to $0.99R_\odot$ and reproduce the NSSL-like feature, especially at low and high latitudes. We argue there that the NSSL is maintained by the radially inward angular momentum transport and the turbulent viscosity on the sheared meridional flow.  \cite{Hotta_2015ApJ...798...51H} fail to reproduce the NSSL in mid-latitude. \cite{matilsky_2019ApJ...871..217M} carry out a similar calculation to \cite{Hotta_2015ApJ...798...51H} and reproduce the NSSL-like feature at high and low latitudes. The authors also fail to reproduce the NSSL in the mid-latitude. They conclude that the detailed construction mechanism of the meridional flow must be understood to reproduce the correct NSSL. In their study, highly rotationally constrained convection called the Busse column, is required to reproduce the solar-like fast equator differential rotation. \cite{Hotta_2015ApJ...798...51H} reduced the solar luminosity and \cite{matilsky_2019ApJ...871..217M} increased the rotation rate in order to enhance the rotational influence on the thermal convection. We note that the decrease in luminosity and the increase in rotation rate have the same effect on the Rossby number. \cite{matilsky_2019ApJ...871..217M} argue that when the rotationally constrained Busse column exists in the deep layer, upflows are rotationally constrained even in the near-surface high Rossby number layer. The efficient generation of the near-surface circulation via the gyroscopic pumping effectively suppresses the construction of the NSSL.\par
  When the previous calculation \citep{Hotta_2015ApJ...798...51H,matilsky_2019ApJ...871..217M} was carried out, we did not have any way to maintain the solar-like DR without using the lowered luminosity, larger rotation rates or enhanced diffusivities (solar convective conundrum). That is, the typical ``high-resolution'' simulations fall into anti-solar differential rotation. \citep{OMara_2016AdSpR..58.1475O,hotta_2023SSRv..219...77H}. \cite{hotta_2021NatAs...5.1100H}(hereafter HK21) and \cite{hotta_2022ApJ...933..199H}(hereafter HKS22) recently provide a possible solution to construct the solar-like differential rotation without using special treatment shown above. We found that the magnetic field plays an important role in the angular momentum transport and construction of the meridional flow. In our calculation, the Busse column is not essential to maintain the fast equator. A strong magnetic field can transport the angular momentum radially outward and the thermal convection can be rotationally less constrained. This process possibly also addresses the NSSL problem since a negative factor for the NSSL, i.e., the Busse column, is not necessary. \par
  \par
  The top boundary of the calculation in HK21 is located at $0.96R_\odot$, and we were not able to reproduce the NSSL. In addition, the short time scale fast convection in the convection zone is a negative factor for the fast equator \citep{Gastine_2013Icar..225..156G}. Thus, we do not know if we can reproduce the solar-like differential rotation even with the layer.\par
  The main purpose of this study is to reproduce the NSSL in an unprecedentedly high-resolution simulation with the top boundary at $r=0.99R_\odot$ and understand the maintenance mechanism of the NSSL.
\section{Model}

We solve the magnetohydrodynamic equation in the spherical geometry $(r,\theta,\phi)$ with R2D2 code \citep{Hotta_2019SciA....eaau2307,hotta_2021NatAs...5.1100H}, where $r$, $\theta$, and $\phi$, are the radius, the colatitude, and the longitude, respectively. The simulation in this study is almost identical to HKS22, but we extend our calculation domain radially and increase the resolution. We adopt the solar rotation rate $\Omega_\odot=2.6\times10^{-6}~\mathrm{s^{-1}}$ for the system rotation rate $\Omega_0$, and the solar luminosity $L_\odot=3.8\times 10^{33}~\mathrm{erg~s^{-1}}$. A non-linear slope-limited diffusion suggested by \cite{Rempel_2014ApJ...789..132R} is adopted. We adopt the standard solar model \citep{Christensen-Dalsgaard_1996Sci...272.1286C} for background stratification. 
Free-slip boundary conditions are used for the velocity at the top and bottom boundaries. The radial velocity is zero at these boundaries. Free boundary condition is adopted for the density and the entropy. The magnetic field is radial and horizontal at the top and bottom boundaries, respectively. We imposed the solar luminosity around the bottom boundary specified in Model S and artificially extract the same amound of the luminosity around the top boundary to drive the thermal convection.
We compare our setting in this study with the High case , i.e., the highest resolution case, in HKS22 in Table \ref{ta:numerical_setting}. We resolved the whole convection zone with 5.4 billion grid points in HKS22. That was the highest resolution calculation to date for covering the whole solar convection zone, but we further increased the number of grid points to 12.9 billion. For analyses, we covert the Yin--Yang grid \citep{Kageyama_2004GGG.....5.9005K} to almost equivalent spherical geometry whose number of grids is $(N_r,N_\theta,N_\phi)=(512\times4096\times8192)$.  We also raise the location of the top boundary $r_\mathrm{max}$ from $0.96R_\odot$ to $0.99R_\odot$. The bottom boundary is at the base of the convection zone $r_\mathrm{min}=0.71R_\odot$. The density contrast $\rho_0(r_\mathrm{min})/\rho_0(r_\mathrm{max})$, which is an essential factor for the convection time scale, is 35.8 in HKS22 and increased to 589 in this study. 
The numbers of scale heights in these studies are 3.57 and 6.37, respectively. 
We note that the subscripts 0 and 1 for thermodynamic variables denote the spherically symmetric adiabatic background value and the perturbation, respectively, in this study.
Since the upper layer introduces short time scale convection as explained in Section \ref{sec:introduction}, the time spacing $\Delta t$ decreases to about 40 s in our setting. Then, the number of time steps is increased from 3 million to 11.5 million in our calculation. The calculation in this study is almost ten times more costly than the High case in HKS22. We continued our calculation for 5250 days. It costs about 75 million CPU hours on the Fugaku super computer.
\par
Table \ref{ta:numerical_setting} also shows the effective diffusivities ($\nu_\mathrm{eff}$, $\eta_\mathrm{eff}$, and $\kappa_\mathrm{eff}$), the volume-averaged RMS velocity $\overline{v}_\mathrm{RMS}$, and some non-dimensional parameters (see the table caption for the definitions). The effective diffusivities are evaluated with mean spherical harmonic degree evaluated with the horizontal kinetic energy spectra at $r=0.83R_\odot$ (see Appendix E of HKS22 for more details). Since we use non-linear artificial diffusion, the evaluation should depend on methods, and the readers should regard the values as references.

\begin{table*}[htbp]
  \centering
  \caption{Comparison of the numerical setting and parameters with the High case in HKS22.
  Non-dimensional parameters are defined as $\mathrm{Re}=\overline{v}_\mathrm{RMS}d/\nu_\mathrm{eff}$, $\mathrm{Ek}=\nu_\mathrm{eff}/(\Omega_0d^2)$, $\mathrm{Ro}=\overline{v}_\mathrm{RMS}/(2\Omega_0 d)$, and
  $\mathrm{Ro_\ell}=\overline{\ell}_\mathrm{u}\overline{v}_\mathrm{RMS}/(\pi \Omega_0 d)$, where $d$ and $\Omega_0=2.60\times10^{-6}~\mathrm{nHz}$ are the depth of the calculation domain and the system rotation rate.
  The mean spherical harmonic degree $\overline{\ell}_\mathrm{u}$ is evaluated with the horizontal kinetic energy $\widehat{E}_\mathrm{h}$ at $r=0.83R_\odot$ as $\overline{\ell}_\mathrm{u}=\int_0^{\ell_\mathrm{max}}\ell \widehat{E}_\mathrm{h}d\ell/\int_0^{\ell_\mathrm{max}} \widehat{E}_\mathrm{h}d\ell$.
  \label{ta:numerical_setting}}
  \begin{tabular}{ccc}
  \hline
  \hline
     & HKS22 (High) & This study \\
   \hline  
  No. of Grids  & $384\times1536\times4608\times2$ & $512\times2048\times\times6144\times2$  \\
  $r_\mathrm{max}$ & $0.96R_\odot$ & $0.99R_\odot$ \\
  $\rho_0(r_\mathrm{min})/\rho_0(r_\mathrm{max})$ & 35.8 & 589 \\
  No. of time steps & $\sim$ 3 million & $\sim$ 11.5 million \\
  $\nu_\mathrm{eff},~\eta_\mathrm{eff},~\kappa_\mathrm{eff}~[\mathrm{cm^2~s^{-1}}]$ & $2.64\times10^{10}$ & $1.69\times10^{10}$ \\
  $\overline{v}_\mathrm{RMS}~[\mathrm{m~s^{-1}}]$ & $108$ & $162$ \\
  Reynolds number ($\mathrm{Re}$) & $7.11\times10^3$ & $1.86\times10^{4}$\\
  Ekman number ($\mathrm{Ek}$)& $3.3\times10^{-5}$ & $1.72\times10^{-5}$ \\
  Rossby number ($\mathrm{Ro}$) & $0.12$ & $0.16$ \\
  Mean spherical harmonic degree ($\tilde{\ell}_\mathrm{u}$) & $202$ & $260$ \\
  Local Rossby number ($\mathrm{Ro}_\mathrm{\ell}$) & $15.4$ & $26.4$ \\
  \hline
  \end{tabular}
\end{table*}

\section{Result}
\subsection{Properties of convection and magnetic field}

\begin{figure*}[htbp]
  \centering
  \includegraphics[width=1.0\textwidth]{\figpath/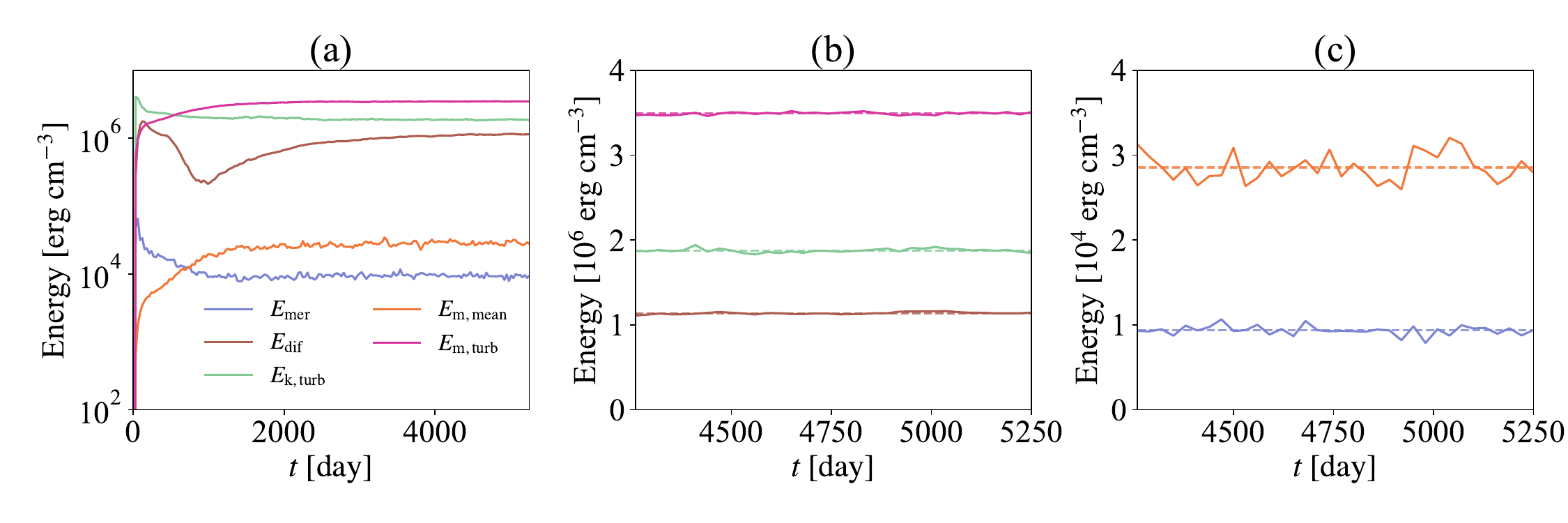}
  \caption{Temporal evolution of energy densities.  Meridional flow (blue), differential rotation (brown)turbulent kinetic (green), mean magnetic (orange), and turbulent magnetic (magenta) energies are shown. The detailed definitions of the energy density are shown in eq. (\ref{eq:energy_density}).
  Panel a shows the overall evolutions of energies. Panels b and c are the focused images in a duration from $t=4260$ to 5250 days. The dashed lines in panels b and c show the averaged energy in the period.
  \label{energy}}
\end{figure*}

\begin{figure*}[htbp]
  \centering
  \includegraphics[width=\textwidth]{\figpath/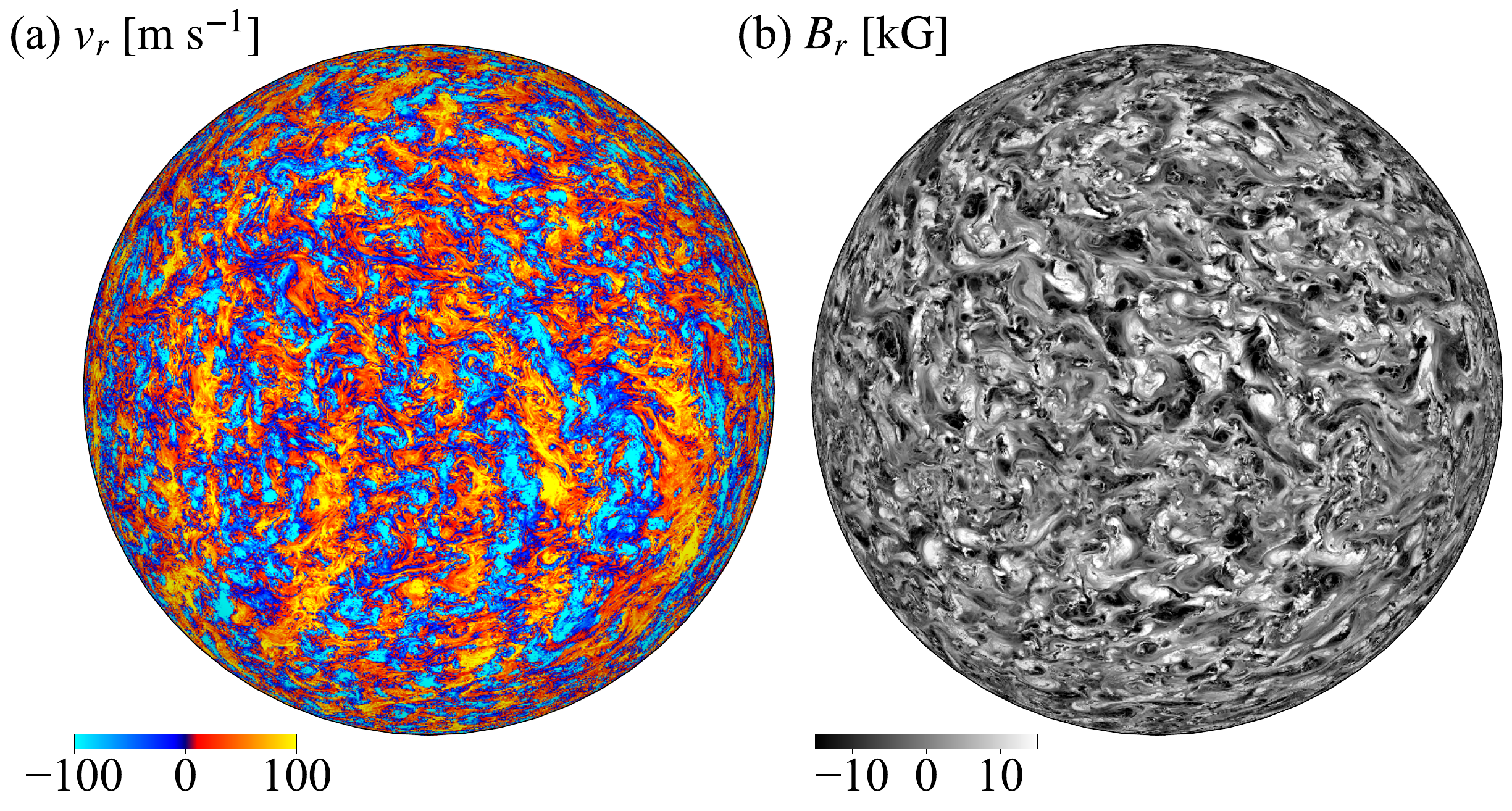}
  \caption{Radial velocity $v_r$ and radial magnetic field $B_r$ at $r=0.95R_\odot$ in $t=5250$ days. A movie is available at \url{https://youtu.be/MbzRv1pRzq4} \label{select_paper}}
\end{figure*}

\begin{figure*}[htbp]
  \centering
  \includegraphics[width=1.0\textwidth]{\figpath/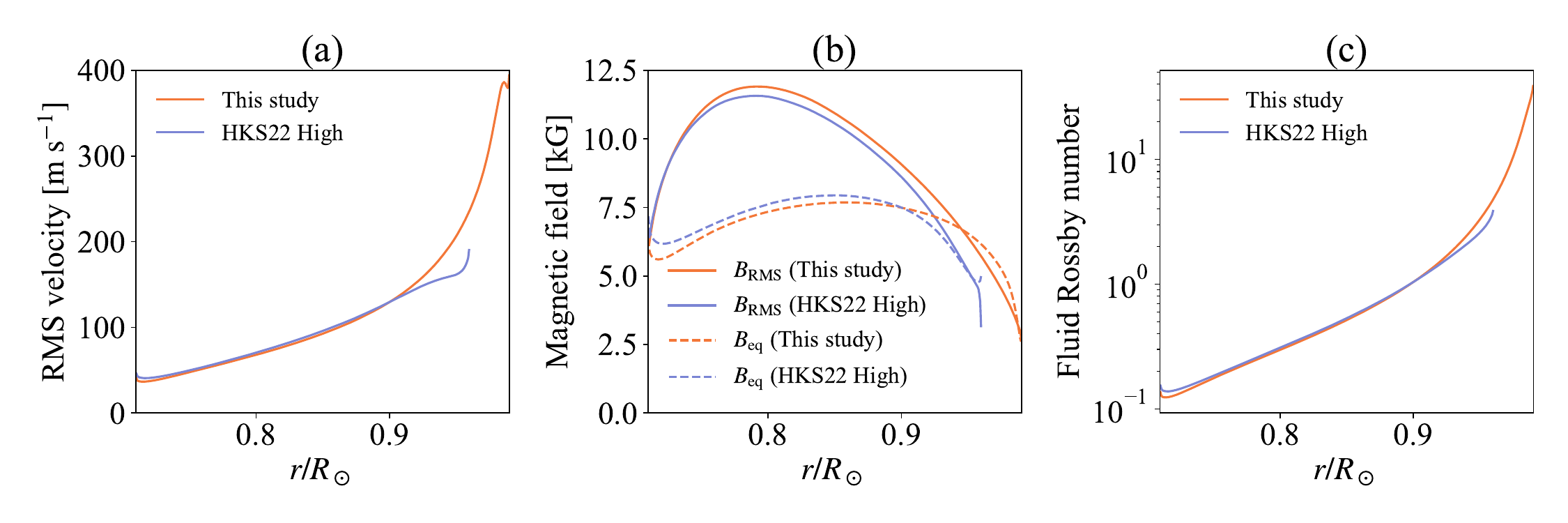}
  \caption{(a) RMS velocity, (b) RMS magnetic field (solid line) and equipartition magnetic field ($\sqrt{4\pi\rho_0}v_\mathrm{RMS}$: dashed line), (c) fluid Rossby number $v_\mathrm{RMS}/(2H_p\Omega_0)$} are shown. The orange and blue lines indicate the result from the High case in HKS22 and in this study, respectively. We use this color format also in the following figures unless otherwise noted.\label{RMS}
\end{figure*}

Fig. \ref{energy} shows the temporal energy evolutions of the meridional flow $E_\mathrm{mer}$, the differential rotation $E_\mathrm{dif}$, turbulent kinetic $E_\mathrm{k,turb}$, mean magnetic $E_\mathrm{m,mean}$, turbulent magnetic $E_\mathrm{m,turb}$, which are defined as
\begin{align}
\begin{aligned}
  E_\mathrm{mer} & = \frac{1}{V}\int_V \frac{1}{2}\rho_0 
  \left(\langle v_r\rangle^2 + \langle v_\theta\rangle^2\right) dV,\\
  E_\mathrm{dif} & =\frac{1}{V}\int_V \frac{1}{2}\rho_0 \langle v_\phi\rangle^2 dV,\\
  E_\mathrm{k,turb} &= \frac{1}{V}\int_V \frac{1}{2}\rho_0 \bm{v}'^2 dV,\\
  E_\mathrm{m,mean} &= \frac{1}{V}\int_V \frac{\langle\bm{B}\rangle^2}{8\pi} dV,\\
  E_\mathrm{m,turb} &= \frac{1}{V}\int_V \frac{\bm{B}'^2}{8\pi} dV,
  \label{eq:energy_density}
\end{aligned}
\end{align}
where $\bm{v}$ and $\bm{B}$ are the fluid velocity and the magnetic field, respectively. $V$ denotes the volume of the whole computational domain. The temporal evolution of the energy densities is similar to HK21 (see their supplementary material). The turbulent magnetic field is the dominant energy over the kinetic energies. All the energies reach a statistically steady state around $t=4200$ days (see Figs. \ref{energy}b and c). Thus, we take temporal averages for the final product, e.g. RMS velocity in Fig. \ref{RMS}a, in the period except for Figs. \ref{energy} and \ref{select_paper}.\par

\begin{figure}[htbp]
  \centering
  \includegraphics[width=0.5\textwidth]{\figpath/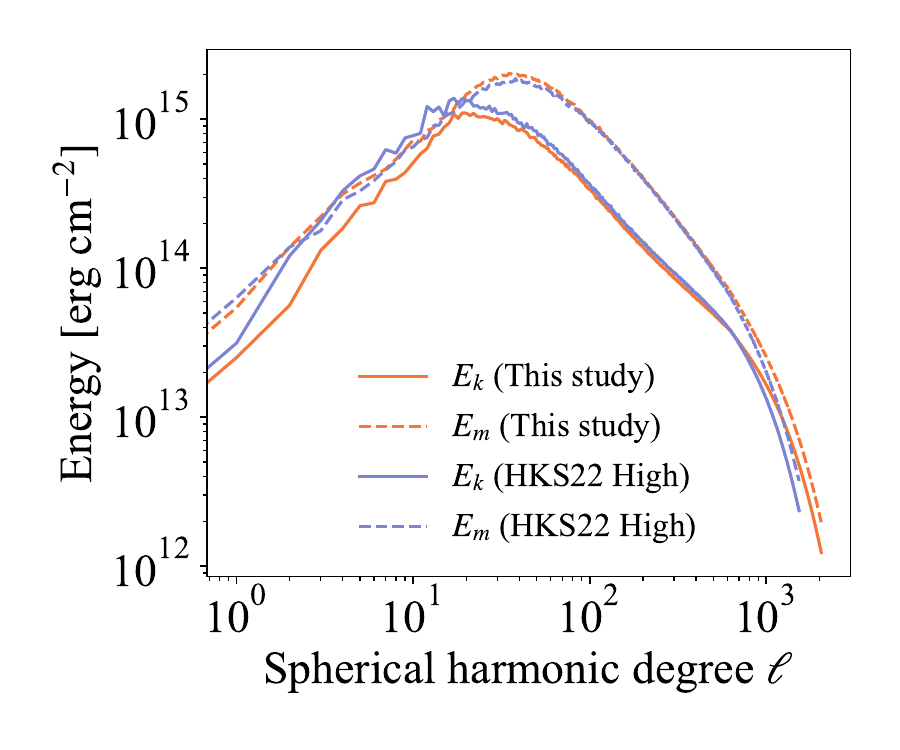}
  \caption{The energy spectra at $r=0.83R_\odot$ are shown. The solid and dashed lines are the kinetic and magnetic energies, respectively.\label{spectra}}
\end{figure}

Fig. \ref{select_paper} shows the radial velocity $v_r$ and the radial magnetic field $B_r$ at $r=0.95R_\odot$. As stated in HKS22, the convection pattern is highly turbulent. The strong magnetic field is mainly concentrated in the downflow region, but the small-scale magnetic field is observed everywhere, even in the upflow region. A movie from $t=0$ to 5250 days is available on \url{https://youtu.be/MbzRv1pRzq4}. The movie nicely describes the temporal evolution of the differential rotation, which has the fast pole at the beginning and the fast equator in the latter phase. $E_\mathrm{dif}$ minimum in Fig. \ref{energy} ($t\sim1000~\mathrm{day}$) roughly corresponds to the transition. The flow is highly turbulent, but we can observe some indication of the Busse column (north-south aligned flows) especially in the latter phase.\par
Fig. \ref{RMS} shows the RMS (root-mean-square) velocity (panel a), the RMS magnetic field, the equipartition magnetic field (panel b) and the fluid Rossby number $v_\mathrm{RMS}/(2H_p\Omega_0)$ (panel c). HKS22 shows that, in their explored resolution, as the resolution increases, the convection velocity $v_\mathrm{RMS}$ decreases and the magnetic field strength $B_\mathrm{RMS}$ increases. When we focus on the deep layer ($r<0.9R_\odot$), this tendency is still noted in this study, but the difference between the orange and the blue lines is tiny. This result may partially suggest that our simulations are close to n umerical convergence. The velocity in the near-surface layer ($r>0.9R_\odot$) is larger than that in HKS22. As discussed in Section \ref{sec:introduction}, the low density and temperature in the near-surface layer ($>0.96R_\odot$), which is not included in HKS22, drives the faster convection (Fig. \ref{RMS}a) and large Rossby numbers ($\sim 30$: Fig. \ref{RMS}c). In HKS22, the magnetic field strength is in a superequipartition state throughout the computational domain. Also, in this study, the magnetic field reaches the equipartition level in deep layer, but the RMS magnetic field strength is smaller than the equipartition magnetic field in the near-surface layer ($>0.945R_\odot$). In addition, for the magnetic field, we have some hope that we could  reached numerical convergence in the deep layer, i.e., the difference between the blue and the orange lines in Fig. \ref{RMS}b is tiny. \par
Fig. \ref{spectra} shows the energy spectra at $r=0.83R_\odot$ defined as,
\begin{align}
  \widehat{E}_\mathrm{kin}(\ell) =& \frac{1}{2} \rho_0 \sum_m \widehat{\bm{v}}\cdot\widehat{\bm{v}}^* \\
  \widehat{E}_\mathrm{mag}(\ell) =& \frac{1}{8\pi} \sum_m \widehat{\bm{B}}\cdot\widehat{\bm{B}}^*,
\end{align}
where $\widehat{Q}$ denotes spherical harmonic transfomred value of quantity $Q$ and $*$ denotes the complex conjugate (see eqs. (14)-(17) in HKS22 for more details).
While the smaller scale flow and magnetic field are introduced, the kinetic and magnetic energy spectra are almost identical to HKS22. We can see a significant difference in the kinetic energy (solid line) on a large scale ($\ell<20$). HKS22 argue that this large-scale suppression is caused by the small-scale turbulence transporting the energy efficiently in higher resolution simulations, while a similar efficient energy transport does not take place on the large scale. The simulation in this study continues the tendency in which the higher resolution suppresses the large-scale convection.


\subsection{Mean flows}
\begin{figure*}[htbp]
  \centering
  \includegraphics[width=0.9\textwidth]{\figpath/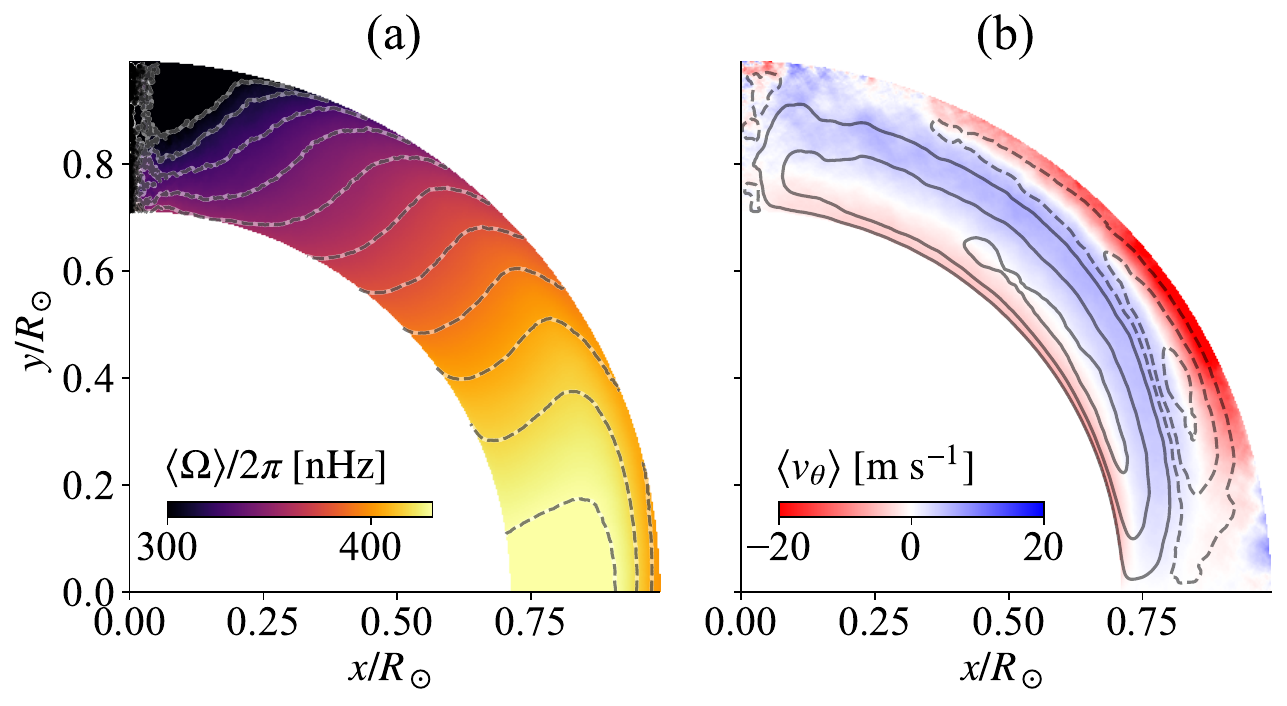}
  \caption{(a) The differential rotation $\langle\Omega\rangle/2\pi$ and (b) the meridional flow $\langle v_\theta\rangle$ are shown. The dashed lines in panel a show the values from 330 to 430 nHz in 10 nHz increments. We note that the system rotation rate $\Omega_0/(2\pi)=413~\mathrm{nHz}$. The black lines in panel b indicate the streamline of the mass flux $\rho_0\langle\bm{v}_m\rangle$. The solid and dashed lines are the clockwise and counter-clockwise flows, respectively.\label{mean_flow}}
\end{figure*}

Fig. \ref{mean_flow} shows the differential rotation $\langle\Omega\rangle/2\pi$ (panel a) and the meridional flow $\langle\bm{v}_\mathrm{m}\rangle$ (panel b), where $\Omega = \Omega_1 + \Omega_0$, $\Omega_1 = v_\phi/(r\sin\theta)$, and $\bm{v}_\mathrm{m}=v_r\bm{e}_r + v_\theta\bm{e}_\theta$. $\bm{e}_r$ and $\bm{e}_\theta$ are the unit vectors in the radial and colatitudinal directions, respectively. Here, we emphasize that we reproduce the solar-like differential rotation even with the near-surface layer. In addition, the NSSL is nicely reproduced in all latitudes. The construction mechanism of the NSSL is the main topic of this study. As for the meridional flow, a prominent poleward flow around the surface consistent with the observation is reproduced \citep{hathaway_1996ApJ...460.1027H}. We note that the poleward flow in HKS22 is weaker than those of the observations. In the middle of the convection zone, the equatorward flow, which is responsible for the equatorward angular momentum transport, exists. We can also observe poleward meridional flow around the base of the convection zone, which is driven by the radially outward magnetic angular momentum transport and the resulting meridional torque (anti-clockwise in the northern hemisphere).

\subsection{Thermal wind balance}
\label{sec:thermal_wind_balance}

\begin{figure*}[htbp]
  \centering
  \includegraphics[width=0.8\textwidth]{\figpath/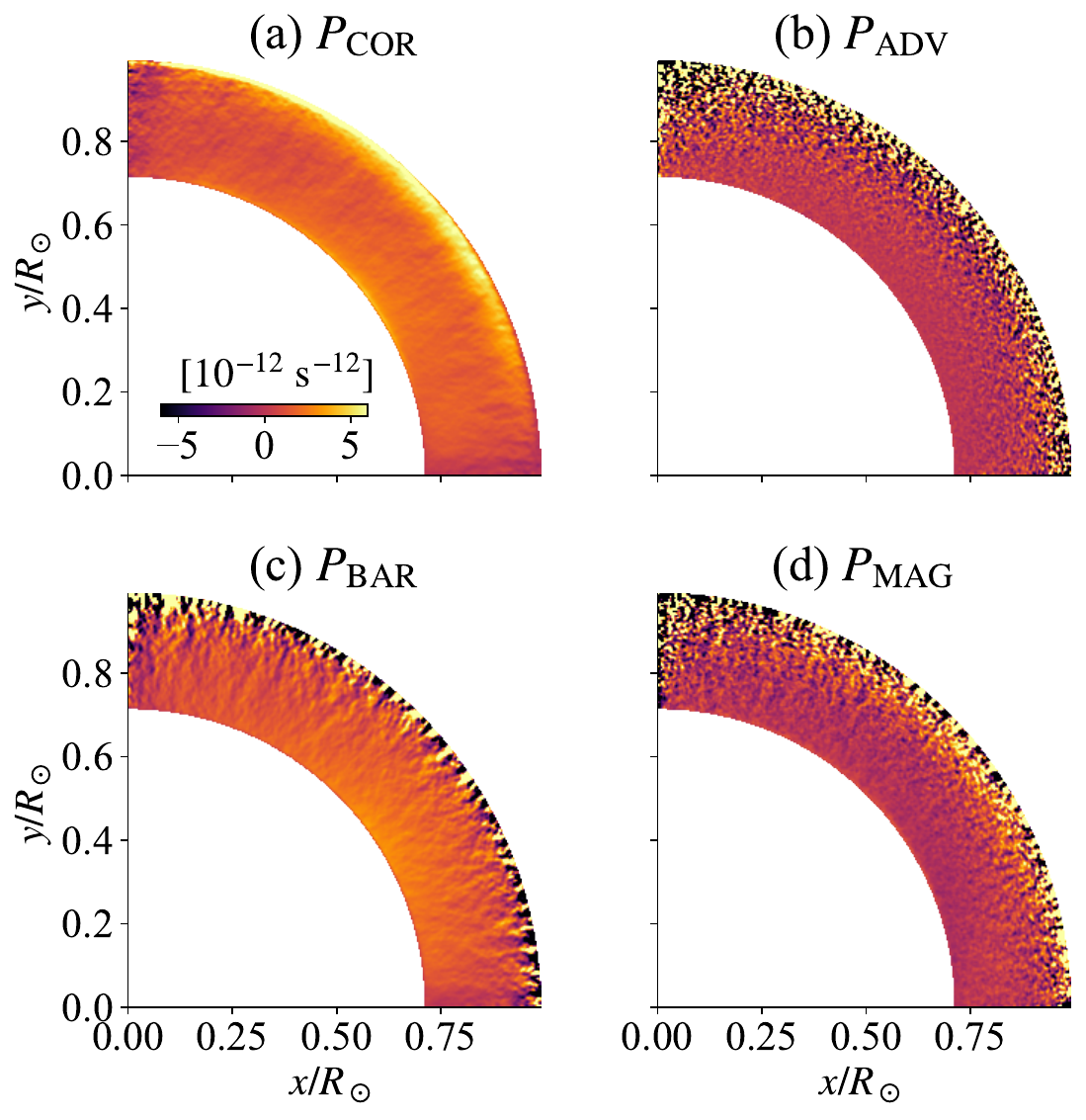}
  \caption{Each term in eq. (\ref{eq:thermal_wind}) is shown on the meridional plane. A Gaussian filter with $5\times5$ grid points width in space without considering the grid size difference is used for decreasing the realization noise.\label{thermal_2d}}
\end{figure*}

\begin{figure*}[htbp]
  \centering
  \includegraphics[width=0.8\textwidth]{\figpath/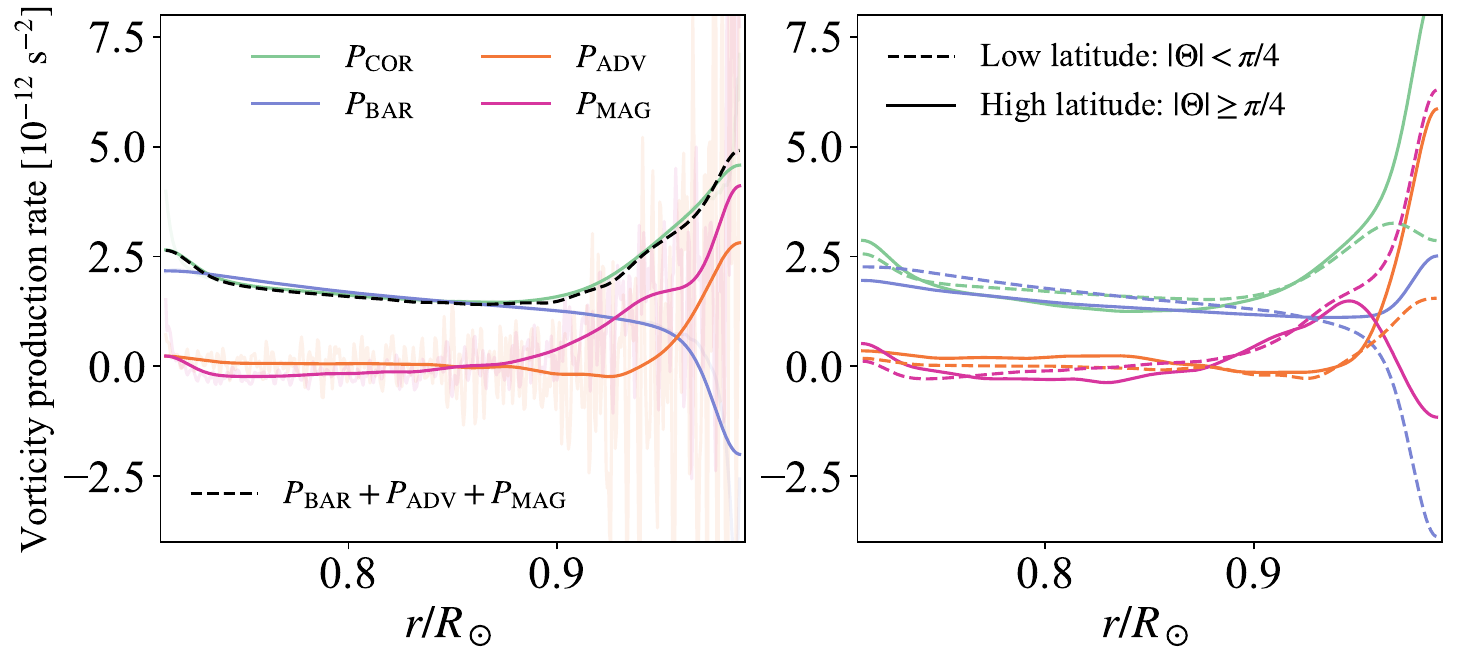}
  \caption{Latitudinally averaged terms in eq. (\ref{thermal_1d}) are shown. Also in this figure, we use a Gaussian filter for averaged data with a 18-grid width in radial direction for decreasing the realization noise. The data before the radial filtering are shown by the transparent lines. To validate the thermal wind balance, we also show $P_\mathrm{BAR}+P_\mathrm{ADV}+P_\mathrm{MAG}$ with the black dashed line. In panel a, the data is averaded over the whole latitude, and in panel b, the data is separated to the low latitude ($|\Theta|<\pi/4$: dashed) and the high latitude ($|\Theta|\geq \pi/4$: solid)\label{thermal_1d}}
\end{figure*}

In this subsection, we discuss the origin of the NSSL in this study. As discussed in Section \ref{sec:introduction}, the NSSL is in part a deviation from the Taylor--Proudman theorem. In order to understand how the deviation from the Taylor-Proudman theorem is maintained in the NSSL, we should analyze the thermal wind balance equation, i.e., the longitudinally averaged longitudinal vorticity equation in steady-state $\partial/\partial t=0$. The thermal wind balance, including the magnetic field, can be written as \cite[see][]{Hotta_2018ApJ...860L..24H}

\begin{align}
  \label{eq:thermal_wind}
  \underbrace{2r\sin\theta\Omega_0\frac{\partial\langle\Omega_1\rangle}{\partial z}}_{P_\mathrm{COR}}
  =&
  \underbrace{\left\langle\nabla\times\left(\bm{v}\times\bm{\omega}\right)\right\rangle_\phi}_{P_\mathrm{ADV}} \nonumber\\
  &+
  \underbrace{\frac{g}{\rho_0 r} \left(\frac{\partial\rho}{\partial s}\right)_p\frac{\partial \langle s_1\rangle }{\partial \theta}}_{P_\mathrm{BAR}}\nonumber \\
  &+
  \underbrace{\left\langle\nabla\times\left[\frac{1}{4\pi\rho_0}\left(\nabla\times\bm{B}\right)\times\bm{B}\right]\right\rangle_\phi}_{P_\mathrm{MAG}},
\end{align}
where we define the vorticity $\bm{\omega}=\nabla\times\bm{v}$, the gravitational acceleration $g$, and the specific entropy $s_1$. We note that $\partial/\partial t = 0$ can be assumed only after the temporal average.
In order to maintain a non-Taylor--Proudman differential rotation $\partial\langle\Omega\rangle/\partial z\neq0$, the Coriolis force term $P_\mathrm{COR}$ must be balanced with the advection term $P_\mathrm{ADV}$, the baroclinic term $P_\mathrm{BAR}$, and/or the magnetic term $P_\mathrm{MAG}$.\par
Fig. \ref{thermal_2d} shows each term in the thermal wind balance equation (eq. (\ref{eq:thermal_wind})). It is clear that the Coriolis force in the middle of the convection zone is maintained by the baroclinic term $P_\mathrm{BAR}$. This is consistent with the previous calculations \cite[][HKS22]{Rempel_2005ApJ...622.1320R,Miesch_2008ApJ...673..557M,Hotta_2018ApJ...860L..24H}. The maintenance of the thermal wind balance for the NSSL is unclear from the figure due to the large fluctuations in $P_\mathrm{ADV}$ and $P_\mathrm{MAG}$ in the layer. Fortunately, the NSSL structure is similar among the different latitudes, and we can latitudinally integrate eq. (\ref{eq:thermal_wind}) to investigate the origin of the NSSL. Fig. \ref{thermal_1d}a shows the averaged result. The black dashed line shows $P_\mathrm{BAR}+P_\mathrm{ADV}+P_\mathrm{MAG}$. While the result has large fluctuation, the green line ($P_\mathrm{COR}$) is comparable to the black dashed line. This indicates that the thermal wind balance with magnetic fields is approximately satisfied in the analyzed period. As can be seen in Fig. \ref{thermal_2d}, the Coriolis force is balanced with the baroclinic term in the deep convection zone ($<0.9R_\odot$). In the NSSL, the baroclinic term works against the maintenance of the NSSL, and the magnetic field and advection have a role in compensating the Coriolis force. We note that the amplitude of $P_\mathrm{BAR}$ does not change between the deep and near-surface layers. Since the importance of $P_\mathrm{ADV}$ and $P_\mathrm{MAG}$ increases due to large Rossby numbers in the near-surface layers, the pure thermal wind balance, i.e., $P_\mathrm{COR}\sim P_\mathrm{BAR}$ does not hold. Fig. \ref{thermal_1d}b show the contributions separately at low and high latitudes. The solid and dashed lines show the low latitude ($|\Theta|<\pi/4$) and the high latitude ($|\Theta|\geq \pi/4$) contributions, respectively, where $\Theta=\pi/2 - \theta$ is the latitude. The magnetic field has a dominant contribution at low latitudes, and the advection term is dominant at high latitudes.

\begin{figure*}[htbp]
  \centering
  \includegraphics[width=0.8\textwidth]{\figpath/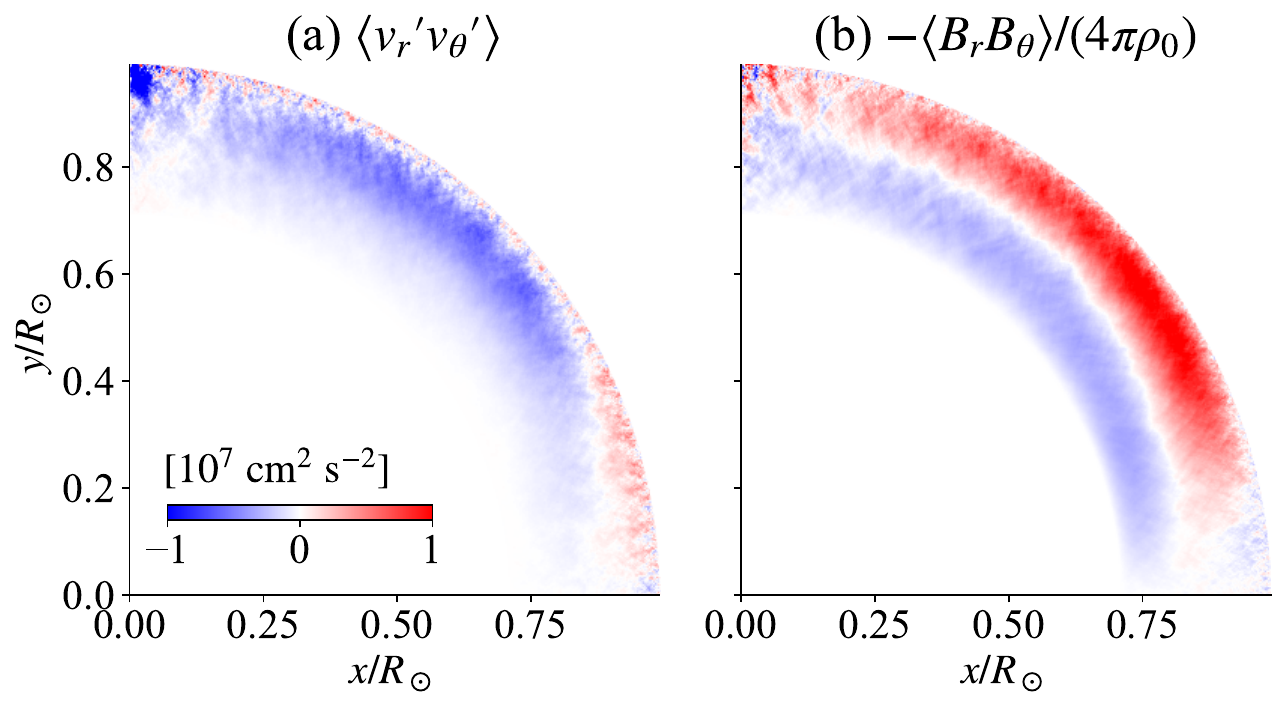}
  \caption{Correlations of (a) $\langle v'_r v'_\theta\rangle$ and (b) $-\langle B_r B_\theta\rangle/4\pi\rho_0$ are shown. \label{meridional_cor}}
\end{figure*}

The origin of the magnetic contribution in the thermal wind balance can be understood with the colatitudinal momentum transport in the radial direction, i.e., temporally averaged $\langle v'_r v'_\theta\rangle$ and $-\langle B_r B_\theta\rangle$. Since the mean magnetic field $\langle\bm{B}\rangle$ is weak (see Fig. \ref{energy}), we do not divide the magnetic field into the mean and perturbed part in this discussion. If $\langle v'_r v'_\theta\rangle$ and/or $-\langle B_r B_\theta\rangle$ are positive in the NSSL, these transport the colatitudinal momentum $\rho v_\theta$ radially outward. The colatitudinal momentum transport accelerates (decelerates) the top (bottom) part of the NSSL in colatitudinal direction, which can be balanced with the Coriolis force. Figs. \ref{meridional_cor}a and b show $\langle v'_r v'_\theta\rangle$ and $-\langle B_r B_\theta\rangle/4\pi\rho_0$, respectively. The result apparently shows that $-\langle B_r B_\theta\rangle/4\pi\rho_0$ has a positive value in the upper half of the convection zone, while the positive $\langle v'_r v'_\theta\rangle$ can be found in high and low latitudes very near the surface.
This result suggests that the correlation of the magnetic field component $\langle B_rB_\theta\rangle$ is negative in the NSSL. We need to understand the mechanism to generate the negative $\langle B_r B_\theta\rangle$. \par

\begin{figure*}[htbp]
  \centering
  \includegraphics[width=0.8\textwidth]{\figpath/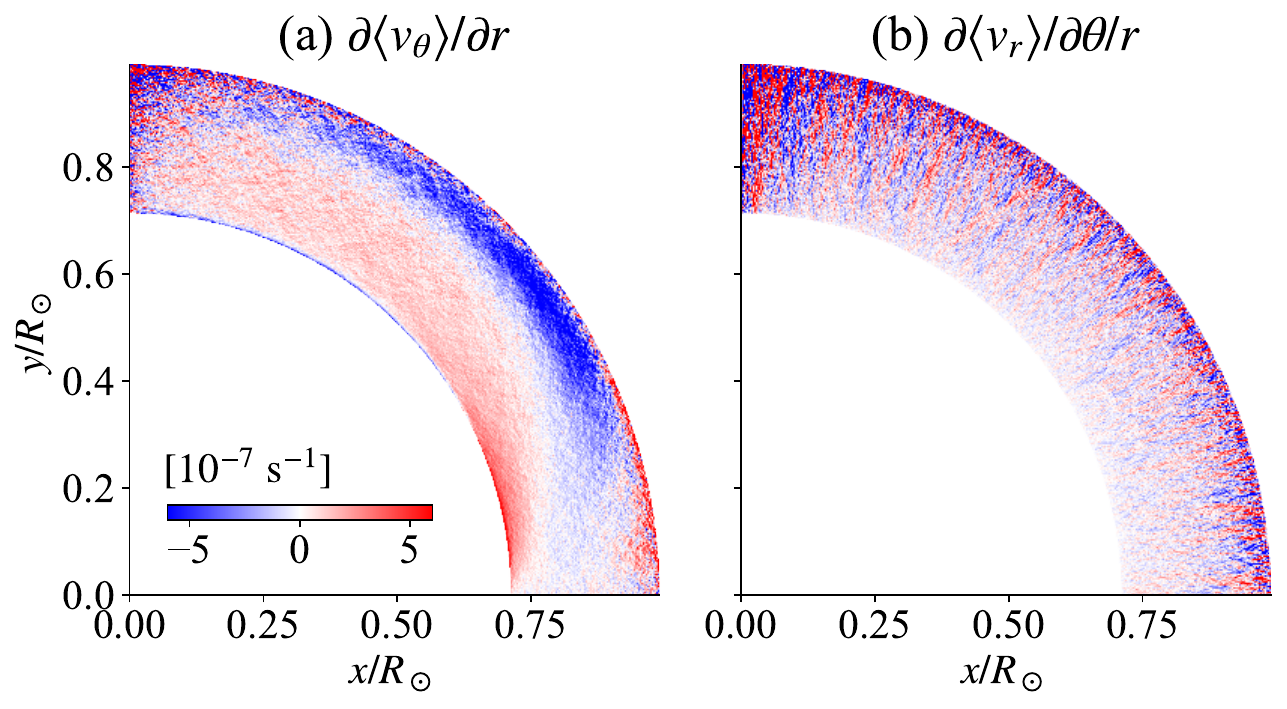}
  \caption{(a) $\partial \langle v_\theta \rangle/\partial r$ and (b) $\partial \langle v_r\rangle/r\partial \theta$ are shown.\label{flow_shear}}
\end{figure*}

A possible origin of the negative correlation  $\langle B_rB_\theta\rangle$ is flow shear.
Since the radial and colatitudinal components of the induction equation are written as
\begin{align}
  \frac{\partial B_r}{\partial t} &= [...] + \frac{B_\theta}{r}\frac{\partial v_r}{\partial \theta}, \\
  \frac{\partial B_\theta}{\partial t} &= [...] + B_r\frac{\partial v_\theta}{\partial r},
\end{align}
the flow shear causes the correlation of the magnetic field components. The sign of the generated correlation is the same as that of the shear. Fig. \ref{flow_shear} shows the flow shear of the meridional flow. The radial gradient of the colatitudinal velocity $\partial \langle v_\theta\rangle/\partial r$ clearly has a negative gradient in the NSSL (Fig. \ref{flow_shear}a). The location of the negative $\partial \langle v_\theta \rangle/\partial r$ is similar to that of negative $\langle B_r B_\theta\rangle$. The result shows that the negative $\langle B_r B_\theta\rangle$ is caused by the shear of the poleward meridional flow.
On the other hand, the colatitudinal gradient of the radial flow $\partial \langle v_r\rangle/\partial \theta /r$ has a messy feature. The radial velocity is not seem to contribute to the correlation generation. \ref{flow_shear}b.
\par

\begin{figure*}[htbp]
  \centering
  \includegraphics[width=\textwidth]{\figpath/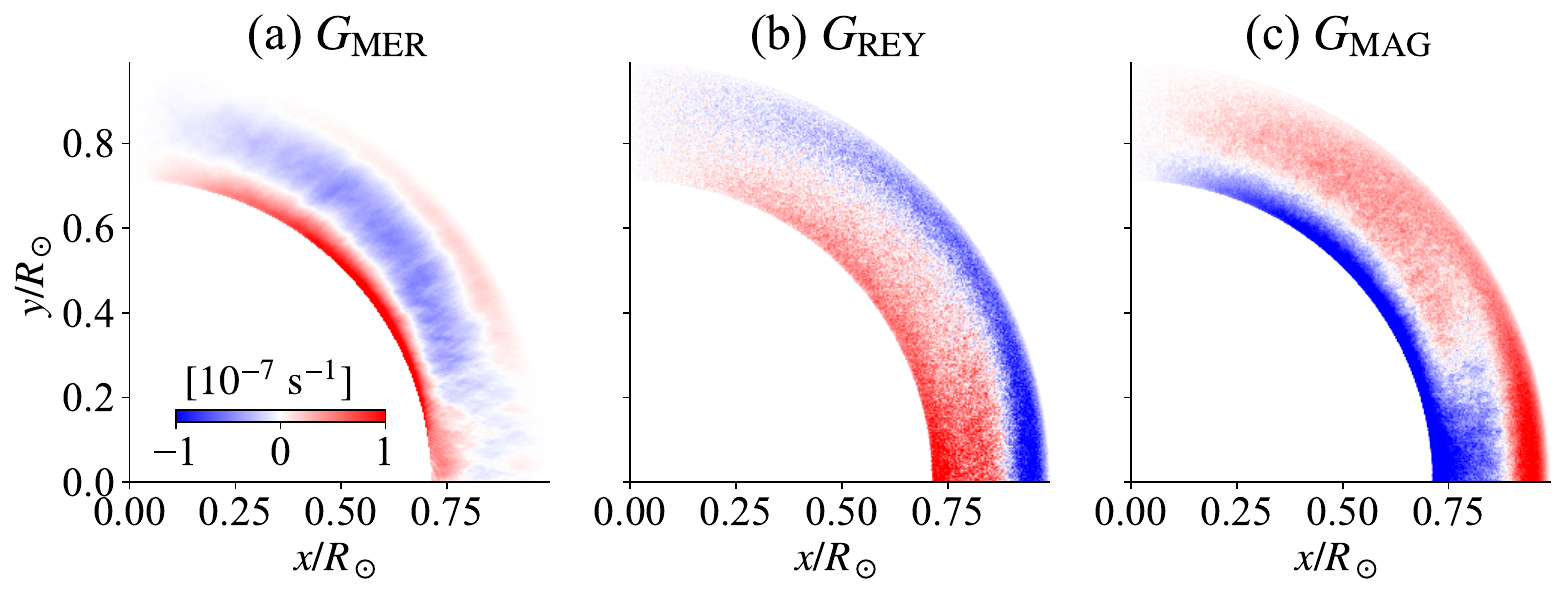}
  \caption{Each term in eq. (\ref{eq:gyroscopic}) is shown.\label{gyroscopic}}
\end{figure*}

\begin{figure*}[htbp]
  \centering
  \includegraphics[width=0.7\textwidth]{\figpath/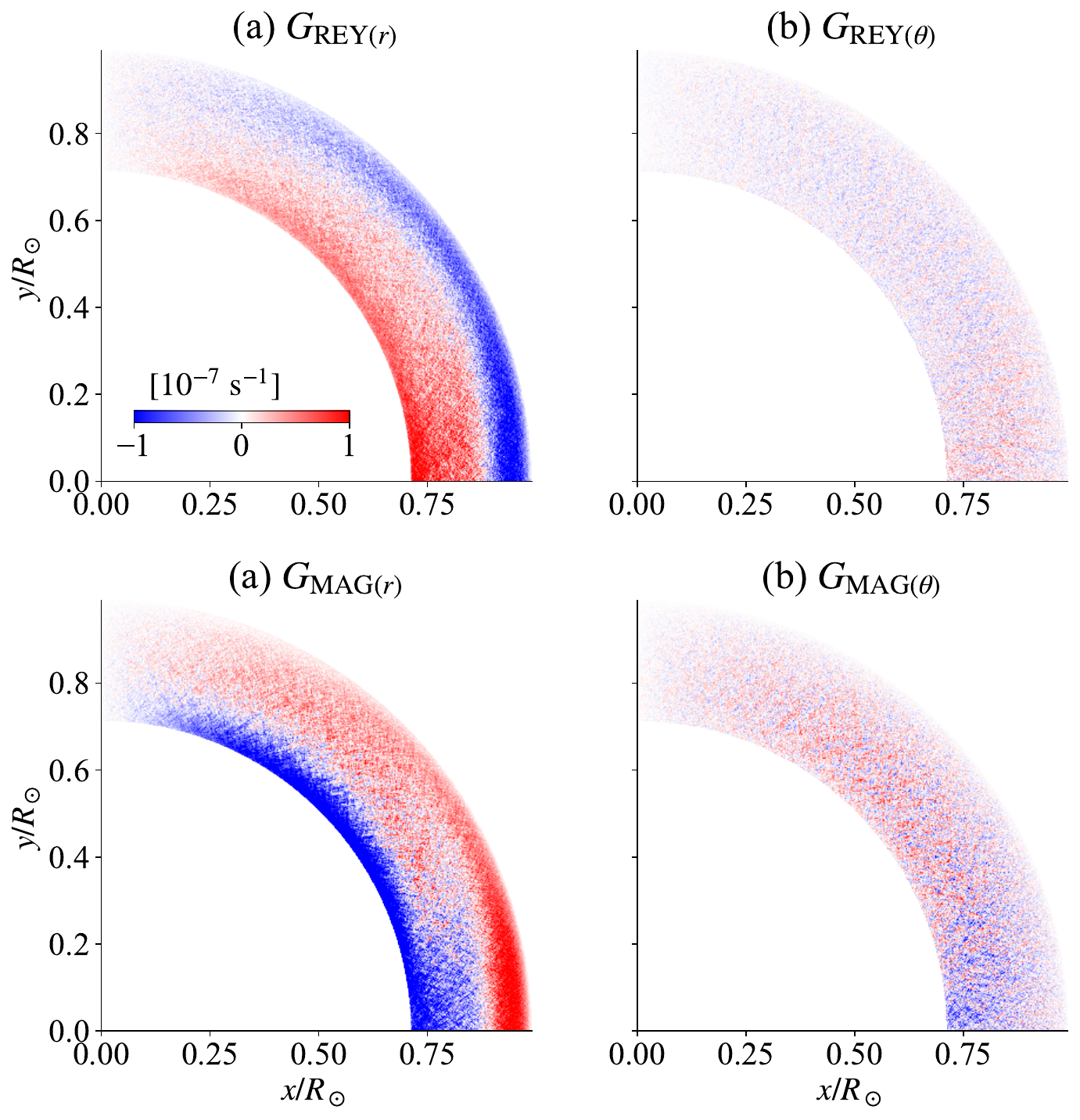}
  \caption{Each term in eq. (\ref{eq:gyroscopic}) is shown.\label{gyroscopic_decompse}}
\end{figure*}

\begin{figure*}
  \centering
  \includegraphics[width=0.6\textwidth]{\figpath/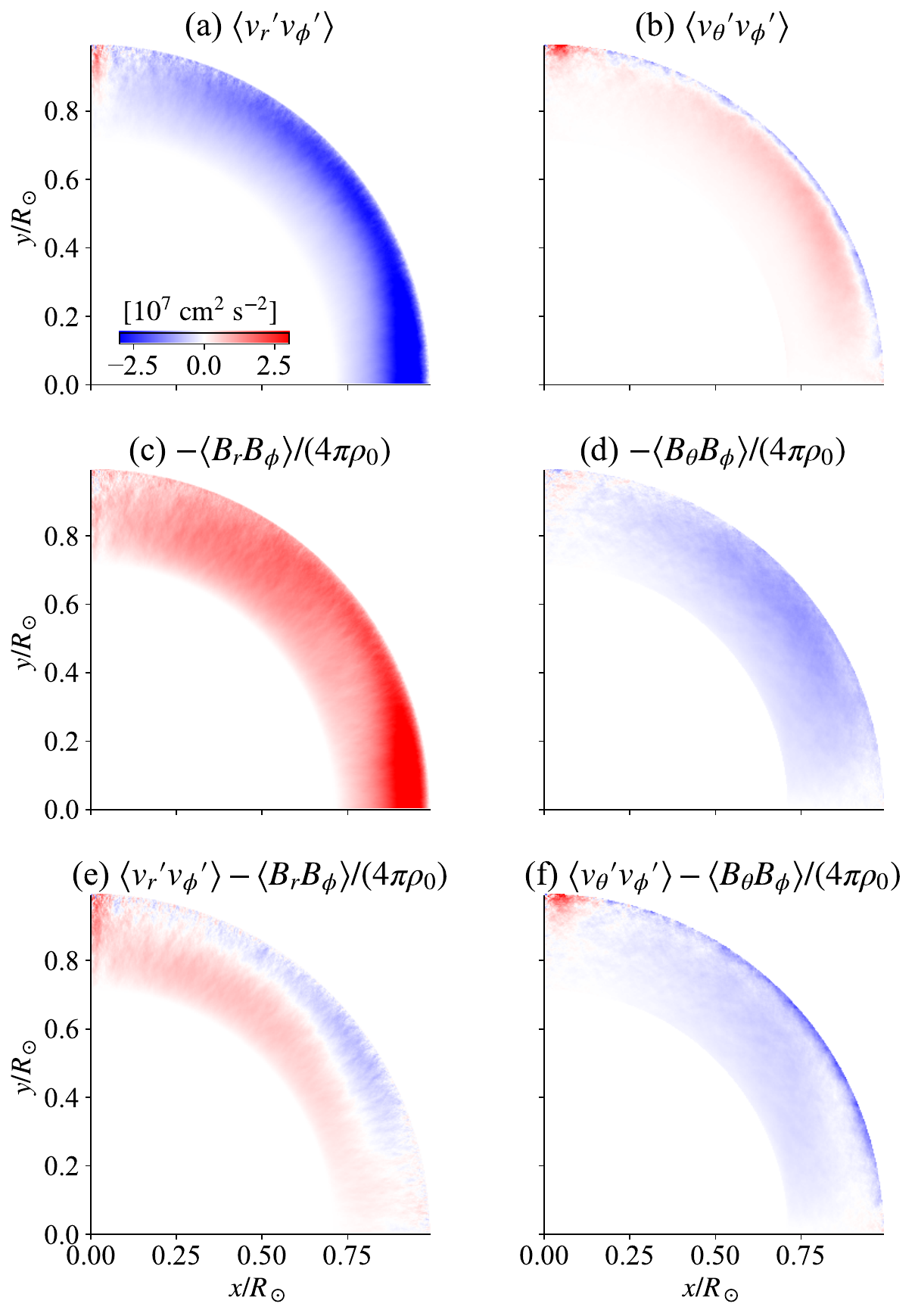}
  \caption{The Reynolds and Maxwell stresses are shown. Panels e and f show the sum of these.\label{angular_momentum_flux}}
\end{figure*}

\subsection{Angular momentum balance}

In this section, we discuss the angular mommentum balance in the NSSL with the gyroscopic pumping. This discussion allows us to understand the mantainance mechanism of the meridinal flow.
The gyroscopic pumping is written as
\begin{align}
  \underbrace{\rho_0\left(\langle\bm{v}_\mathrm{m}\rangle\cdot\nabla\right)
  \langle\mathcal{L}\rangle}_{-G_\mathrm{MER}}
  =& \underbrace{-\nabla\cdot\left(\rho_0\lambda \langle \bm{v}'_\mathrm{m} v'_\phi\rangle\right)}_{G_\mathrm{REY}} \nonumber\\
  &  \underbrace{-\nabla\cdot\left(\frac{\langle \bm{B}_\mathrm{m}B_\phi\rangle}{4\pi}\right)}_{G_\mathrm{MAG}}.
  \label{eq:gyroscopic}
\end{align}
where $\lambda=r\sin\theta$ and shows the direction perpendicular to the rotational axis. 
We again note that we need a temporal average to validate the equation.
Since the specific angular momentum $\mathcal{L}=\lambda^2\Omega\sim\lambda^2 \Omega_0$ mainly depends on $\lambda$ \citep[see][]{Miesch_2011ApJ...743...79M}, the angular momentum transport by the meridional flow is approximated as
\begin{align}
  G_\mathrm{MER} = - \rho_0\left(\langle\bm{v}_\mathrm{m}\rangle\cdot\nabla\right)
  \sim - \rho_0 v_\lambda\frac{\partial \langle\mathcal{L}\rangle}{d\lambda}.
\end{align}
Since $\partial\langle\mathcal{L}\rangle/\partial \lambda>0$ in the solar convection zone, the poleward flow ($v_\lambda<0$) leads to an increase of the angular momentum, i.e., $-\rho_0\left(\langle\bm{v}_\mathrm{m}\rangle\cdot\nabla\right)>0$. Thus, the origin of the poleward meridional flow in the NSSL is the torque, which decreases the angular momentum.
Fig. \ref{gyroscopic} shows each term in eq. (\ref{eq:gyroscopic}). We note that Fig. \ref{gyroscopic}a shows $G_\mathrm{MER}$, while $-G_\mathrm{MER}$ is provided on the left side of eq. (\ref{eq:gyroscopic}). 
As explained, the poleward meridional flow increases the angular momentum in the NSSL. Fig. \ref{gyroscopic}b clearly shows that the Reynolds stress torque decreases the angular momentum and compensates for the contribution by the meridional flow. Thus, the driving mechanism of the meridional flow is the turbulence.\par

We decompose $G_\mathrm{REY}$ and $G_\mathrm{MAG}$ to the radial and colatitudinal contributions as:
\begin{align}
  G_\mathrm{REY} =&
  \underbrace{-\frac{1}{r^2}\frac{\partial}{\partial r}\left(r^2 \rho_0\lambda \langle v'_r v'_\phi\rangle\right)}_{G_{\mathrm{REY}(r)}} \nonumber\\
  & \underbrace{-\frac{1}{r\sin\theta}\frac{\partial}{\partial \theta}\left(\sin\theta \rho_0\lambda \langle v'_\theta v'_\phi\rangle\right)}_{G_{\mathrm{REY}(\theta)}}, 
  \label{eq:gyroscopic_decomposed_rey}\\
  G_\mathrm{MAG} =&
  \underbrace{\frac{1}{r^2}\frac{\partial}{\partial r}\left(r^2 \lambda \frac{\langle B_r B_\phi\rangle}{4\pi}\right)}_{G_{\mathrm{MAG}(r)}} \nonumber\\
  & \underbrace{+\frac{1}{r\sin\theta}\frac{\partial}{\partial \theta}\left(\sin\theta \lambda \frac{\langle B_\theta B_\phi\rangle}{4\pi}\right)}_{G_{\mathrm{MAG}(\theta)}}.
  \label{eq:gyroscopic_decomposed_mag}
\end{align}
Fig. \ref{gyroscopic_decompse} show the decomposed torque. The radial contribution is dominant both for Reynolds and Maxwell stresses. 
Fig. \ref{angular_momentum_flux} shows the Reynolds and Maxwell stresses. 
As shown in Fig. \ref{gyroscopic_decompse} the radial angular momentum transport is dominant.
The Reynolds stress always transports the angular momentum radially inward, and the Maxwell stress does so in the opposite direction. This inward angular momentum transport is in part the reason why we have the poleward meridional flow. There is a competition between the inward transport by the Reynolds stress and the outward transport by the Maxwell stress. The sum of these is shown in Fig. \ref{angular_momentum_flux}e. In the deep layer ($<0.9R_\odot$), the magnetic field is dominant, and the outward angular momentum transport remains. This transport leads to the clockwise meridional flow (see Fig. \ref{mean_flow}b). The meridional flow is responsible for the fast equator  (see HKS22). In the upper layer ($>0.9R_\odot$), the flow becomes dominant and the radially inward transport remains. 
This drives the poleward meridional flow around the solar surface.\par
As shown in Section \ref{sec:thermal_wind_balance}, the shear of the meridional flow is the critical factor to maintain the NSSL. We can argue that the meridional flow is sheared due to the sheared Reynolds stress $G_\mathrm{REY}$. In this study, we cannot reasonably explain the origin of shear in $G_\mathrm{REY}$, but the large density contrast in the near-surface layer may be a key.

\section{Summary and Discussion}
\label{sec:summary}
\begin{figure*}[htbp]
  \centering
  \includegraphics[width=0.7\textwidth]{\figpath/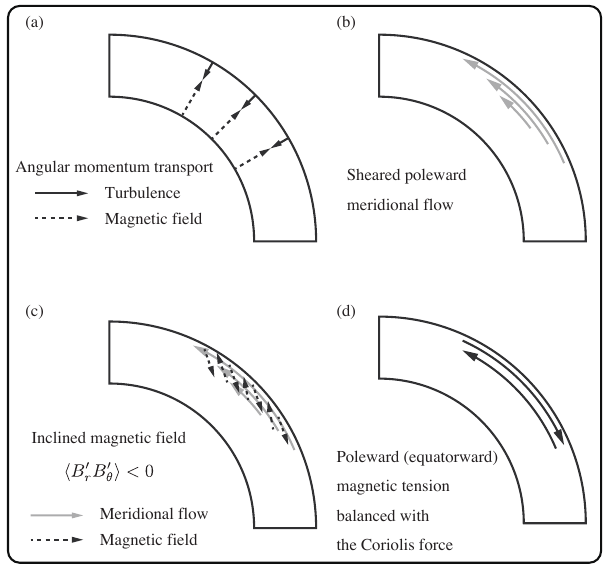}
  \caption{The obtained mechanism revealed in this study is shown.\label{summary}}
\end{figure*}

We carry out an unprecedentedly high-resolution simulation for the solar convection zone. The fast equator, the poleward meridional flow around the top boundary, and the NSSL  are reproduced simultaneously. The overall mechanism related to the newly discovered magnetic role is described in Fig. \ref{summary}.
(a) While the angular momentum transport is radially outward in the deep convection zone due to the dominance of the magnetic field, the turbulence leads to transport in the opposite direction in the near-surface layer.
(b) The radially inward transport decreases (increases) the angular momentum in the top (bottom) of the near-surface layer. This process causes the poleward meridional flow around the surface. In our calculation, the poleward flow is sheared, i.e., increses in radial direction.
(c) The turbulent magnetic field is stretched by the sheared meridional flow and has a negative correlation $\langle B_r B_\theta\rangle<0$.
(d) The magnetic tension (Maxwell stress) caused by the sheared meridional flow is balanced by the Coriolis force in the NSSL. Thus, the NSSL is maintained.\par
\cite{Hotta_2015ApJ...798...51H} suggest that the turbulent viscosity is balanced with the Coriolis force and maintains the NSSL. The reproduced NSSL in \cite{Hotta_2015ApJ...798...51H} is weak compared with observations. The turbulent viscosity also has role to maintaining the NSSL in this study especially in high latitude very near surface (Fig. \ref{thermal_1d}b). The conbination of the turbulent viscosity and the magnetic field can maintain the solar-like NSSL in the simulation. In this study, the magnetic field plays almost the same role as the turbulent viscosity in \cite{Hotta_2015ApJ...798...51H}. The magnetic field energy is comparable to the kinetic energy even in the NSSL, and the magnetic field breaking the thermal wind balance and constructs the NSSL to be more consistent with the observations. In the previous model, we needed to have a low Rossby number regime in the deep convection zone to have the fast equator in addition to the high Rossby number regime in the near-surface layer. 
\begin{figure}[htbp]
  \centering
  \includegraphics[width=0.45\textwidth]{\figpath/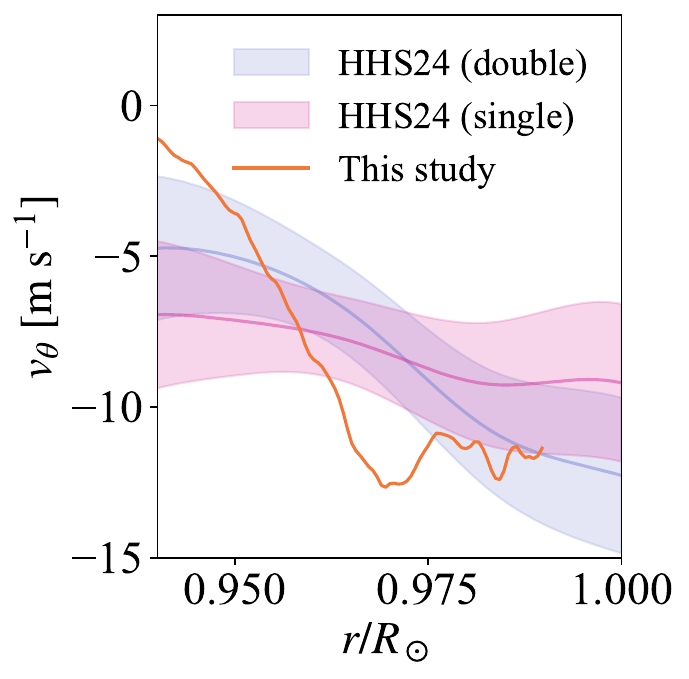}
  \caption{Comparison of the meridional flow in this study and helioseismic results are shown. All data show the colatitudinal meridional flow at $\theta=\pi/4$. The orange line indicates the result in this study. The blue and red lines show the result with helioseismic inversion \citep{hatta_2024ApJ...972...79H}. The shading shows the $1\sigma$ uncertainty. The blue and red lines correspond to double and single cell meridional flow in a hemisphere.  \label{observe_comp}}
\end{figure}
In this study, the local Rossby number shown in Table \ref{ta:numerical_setting} is not low, and the fast equator is constructed by the magnetic field. Thus, the Busse column is less prominent, and we can nicely construct the NSSL. \par
Several recent studies have investigated the possibility that the NSSL can be explained by thermal wind balance alone, without invoking fully nonlinear or turbulent effects. \cite{choudhuri_2021SoPh..296...37C} and \cite{jha_2021MNRAS.506.2189J} demonstrated that a convincing NSSL structure can emerge from a simplified mean-field model that satisfies thermal wind balance, even in the absence of turbulence. These models provide valuable insight into the dynamical consistency of the NSSL and suggest that baroclinicity alone may be sufficient under certain conditions.
\cite{matilsky_2023MNRAS.526L.100M}, on the other hand, constructed a semi-analytic model of the stellar thermal wind as a consequence of baroclinic oblateness and investigated the resulting differential rotation. He argued that the classical thermal wind balance, particularly the baroclinic term, fails to account for the observed NSSL structure. Instead, he concluded that the NSSL is not in thermal wind balance but rather in a force balance involving the centrifugal force, gravity, and pressure gradient (the so-called "GPR balance").
Our simulation results offer a complementary perspective: we confirm that the classical thermal wind balance (i.e., the balance between the Coriolis and baroclinic terms) does not hold in the NSSL. Instead, the Coriolis term is primarily balanced by nonlinear advection and magnetic tension (see Figs \ref{thermal_2d} and \ref{thermal_1d}), which become increasingly important at large Rossby numbers near the surface. This indicates that fully nonlinear and magnetohydrodynamic effects are essential for maintaining the NSSL in our model. Thus, while the earlier studies suggest that thermal wind balance may be sufficient in idealized or laminar conditions, our results highlight the crucial role of turbulence and magnetic fields in sustaining the solar-like NSSL in a realistic dynamical regime.
Our numerical simulation may be reaching numerical convergence. The large-scale structure, except for the NSSL, is almost consistent with our previous simulations (HKS22). Remaining important issue is validation of our numerical simulation with observations. We compare our result with helioseismology for the meridional flow \citep[][hereafter HHS24]{hatta_2024ApJ...972...79H}. HHS24 use \cite{gizon_2020Sci...368.1469G} travel time data to invert the meridional flow. They use different assumptions for the inversion to make both single and double cell meridional flow possible. The shear of the meridional flow in this study is stronger than those of helioseismic results. There is shears in the meridional flow in most of helioseimic inversion results \citep[e.g.,][]{zhao_2013ApJ...774L..29Z,rajaguru_2015ApJ...813..114R,jakiewicz_2015ApJ...805..133J}, but these are weaker than those in this study. The shear of the meridional flow is an essential factor in maintaining the NSSL. Since the inversion result significantly depends on assumptions, more detailed comparisons between simulations and helioseismic results are planned to further investigate the validity of our simulation in the future publication.
\par

\begin{acknowledgments}
  H.H. is supported by JSPS KAKENHI grants No. JP20K14510, JP21H04492, JP21H01124, JP21H04497, JP23H01210, and MEXT as a Programme for Promoting Researches on the Supercomputer Fugaku (JPMXP1020230504). The author appreciates Dr. Yoshiki Hatta for providing the meridional flow data from their helioseismic inversion. The results were obtained using the Supercomputer Fugaku provided by the RIKEN Center for Computational Science (hp220173, hp230204, hp230201, hp240219, and hp240212).
\end{acknowledgments}

\software{R2D2 \cite{Hotta_2019SciA....eaau2307,Hotta_2020MNRAS.494.2523H,hotta_2021NatAs...5.1100H}}


\bibliographystyle{aasjournal}

\end{document}